# THIRD COMPONENTS WITH ELLIPTICAL ORBITS IN THE ECLIPSING BINARIES: BD AND, SV Cam, V0836 CYG & XZ CMI

Tvardovskyi D.E.

[1] Odessa I. I. Mechnikov National University, Odessa, Ukraine

[2] Department "Mathematics, Physics and Astronomy", Odessa National Maritime University, Odessa, Ukraine

**Abstract.** This is our regular research in the field of cyclic O-C changes and third components. In this article the results of four new stars investigation is described. These stars are: BD And, SV Cam, V0836 Cyg and XZ CMi. All of them have cyclic O-C curve with linear (BD And, SV Cam and XZ CMi) or parabolic trend (V0836 Cyg). We computed the mass transfer rate, minima possible mass of the third component and their errors for each of the researched stars.

All of the researched stars are well-known eclipsing binaries which were observed during long period of time. Thus, a lot of photometric, photoelectric observations were done by amateur astronomers and by specialized telescopes. All available data from databases AAVSO [1] and BRNO [2] was used in this research as well as results of the previous investigations made by other authors. Firstly, we took some important general parameters from General Catalogue of Variable Stars (GCVS, [3]) and other researches.

Table 1. Some parameters of the studied eclipsing binaries

| Stellar systems | Initial epoch (JD-2400000) | Period (days) | $M_1, M_\odot$ | $M_2, M_\odot$ | Reference |
|---|---|---|---|---|---|
| BD And | 45253.417 | 2.043926 | 1.145±0.053 | 1.004±0.047 | [4] |
| SV Cam | 52500.3873 | 0.4629057 | 1.47±0.06 | 0.87±0.06 | [5] |
| V0836 Cyg | 52500.1133 | 0.593072 | 1.29±0.07 | 0.57±0.03 | [6] |
| XZ CMi | 44853.4903 | 0.6534122 | 1.7 | 0.7 | [7] |

For XZ CMi errors of the masses were not computed. Thus, they were estimated as 7% of the stellar masses, because it is average value of the errors.

Secondly, all previous articles and abstracts were analyzed. Here is a brief overview of previously published results together with our O-C curves.

Table 2. General description of the most important results from publications of other authors.

| Stellar system | BRNO points | AAVSO points | 3rd component's mass | Orbital elements |
|---|---|---|---|---|
| BD And | 170 | 44 | [4], [5] | [5] |
| SV Cam | 1575 | 153 | [8], [9], [10], [11], [12], [13] | [6], [7], [8], [9], [10], [11], [12], [13] |
| V0836 Cyg | 210 | 20 | - | - |
| XZ CMi | 184 | 18 | [14] | [14], [15] |

On all figures pink dots are BRNO observations, blue ones are moments of minima which were computed using AAVSO data. Black line is approximation, in addition the ±σ and ±2σ confidence intervals are shown, where σ is an unbiased estimate or the r.m.s. deviation of the points from the fit.

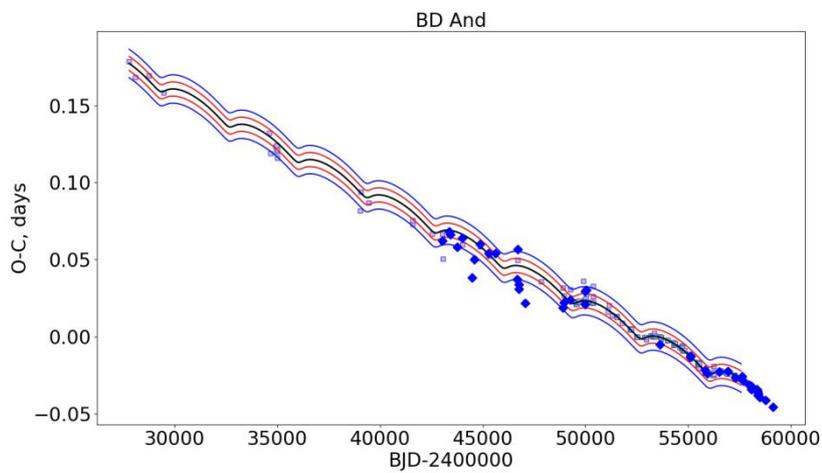

Fig. 1 O-C curve of BD And

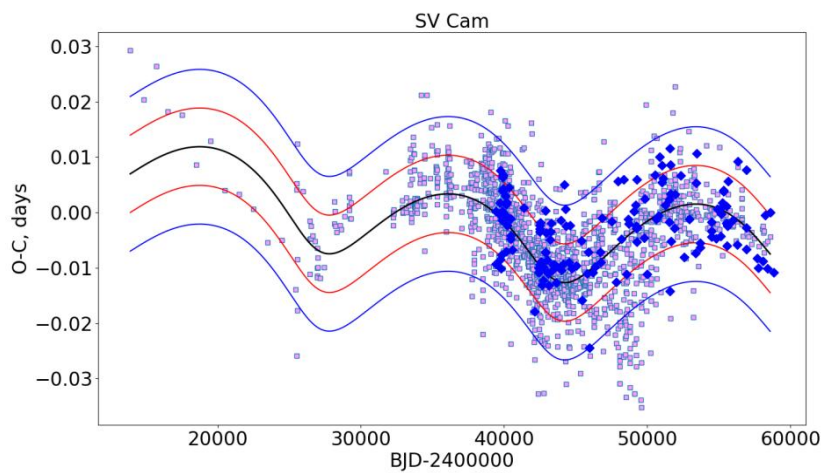

Fig. 2 O-C curve of SV Cam

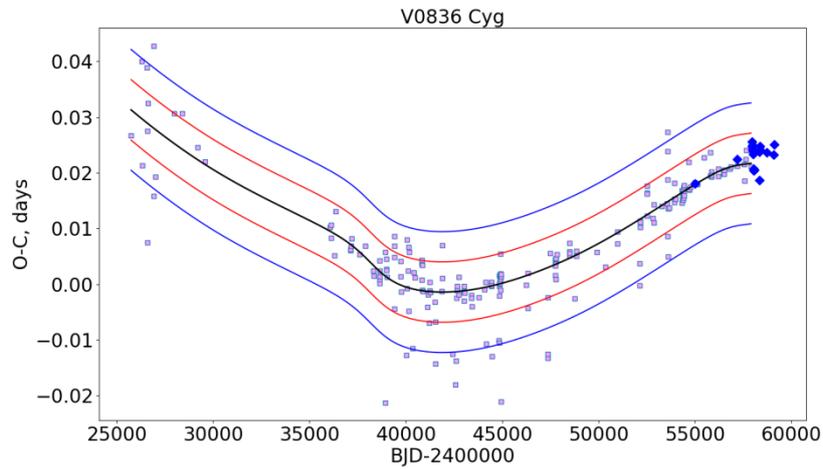

Fig. 3 O-C curve of V0836 Cyg

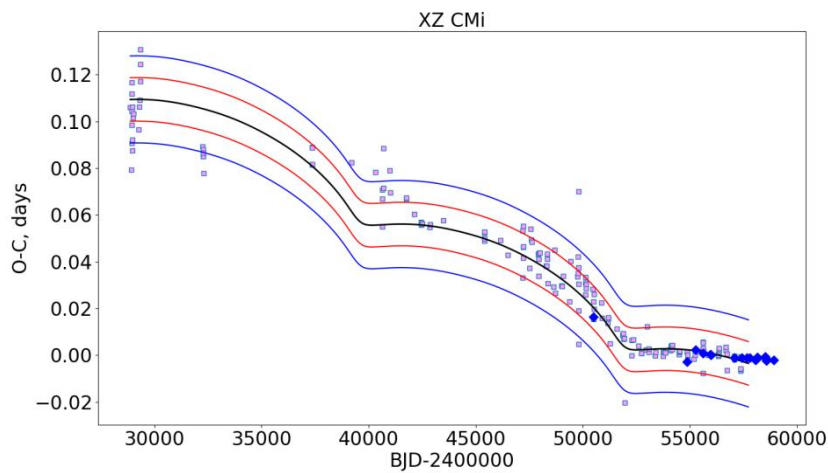

Fig. 4 O-C curve of XZ CMi

Now it is necessary to describe general aspects of the processing algorithm:

1. Collecting data from database BRNO;
2. Downloading observations from AAVSO;
3. Splitting AAVSO data onto separate minima;
4. Obtaining moment of extremum for each minimum;
5. Joining data form BRNO and obtaining moments of minima;
6. Obtaining values of O-C;
7. Plotting and approximating O-C curves;
8. Obtaining period of cyclic and rate of stable O-C changes;
9. Computing parameters of the physical processes that cause such changes.

For calculating moments of minima from AAVSO observations the software MAVKA was actively used. This code was kindly provided by K.D. Andrych and I.L. Andronov [16], [17], [18]. As the result, 235 minima were obtained.

Discussion of the used methods was published in earlier investigation of Odessa variable stars researchers group. Contrary to studies of observations near extrema, in intermediate polars, there are two periods (the orbital period and the spin period of the magnetic white dwarf, sometimes with a dominating double frequency). In this case, two-periodic (multi-harmonic, if needed) approximations are used ([19], [20], [21]). The presence of third bodies around cataclysmic binary systems may also be suggested in [22]. To study night-to-night variability and variations of phases was done using the trigonometric polynomial of the 4-th order [23]. Similarly, for the statistically optimal approximation, for pulsating variables, the best fit order of the trigonometric polynomial is used in [24]-[34].

Table 3. Results of calculations and O-C approximation parameters.

| Value | BD And | SV Cam | V0836 Cyg | XZ CMi |
|---|---|---|---|---|
| $\alpha, 10^{-12} \frac{1}{days}$ | - | 11.6±2.3 | 100.5±4.8 | - |
| $\beta, 10^{-6}$ | -6.87±0.02 | -1.15±0.21 | -8.72±0.43 | -4.33±0.03 |
| $\gamma, days$ | 0.362±0.001 | 0.021±0.005 | 0.190±0.009 | 0.232±0.002 |
| $a \sin i, 10^6\ km$ | 111±3 | 188±3 | 101±7 | 256±8 |
| $e, 1$ | 0.586±0.046 | 0.317±0.019 | 0.610±0.046 | -0.686±0.035 |
| $\omega, rad$ | 5.193±0.047 | 4.232±0.051 | 0.126±0.074 | 1.121±0.078 |
| $t_0, MJD$ | 3338±9 | 6588±136 | 9128±334 | 3615±112 |
| $T, days$ | 9347±123 | 10052±286 | 10330±987 | 8711±432 |
| $\dot{M}, 10^{-9} \frac{M_\odot}{year}$ | - | 5.1±1.4 | 22.6±1.8 | - |
| $M_3, M_\odot$ | 0.118±0.009 | 0.199±0.073 | 0.135±0.032 | 0.299±0.020 |

**Acknowledgements**

This research was done as the part of project was done as the part of the projects Inter-Longitude Astronomy [35], [36], UkrVO [37], [38] and AstroInformatics [39] as well as previous researches [40], [41], [42].

We sincerely thank to Ivan L. Andronov for fruitful discussions and to AAVSO and BRNO databases for providing data for this research. In addition, we are grateful to Kateryna D. Andrych and Ivan L. Andronov [16], [17], [18], [43], [44], [45] for providing software MAVKA that made this investigation possible.